\documentclass[reprint,showpacs,preprintnumbers,amsmath,amssymb,a4paper,nofootinbib,superscriptaddress,floatfix,longbibliography,showkeys, twocolumn]{revtex4-1}

\usepackage{lineno}
\usepackage{blindtext, color}

\usepackage{graphicx}
\graphicspath{{./figs/}}

\usepackage{dcolumn}
\usepackage{bm}
\usepackage{amsmath}
\usepackage{todonotes}
\usepackage{placeins} 

\usepackage{color}
\usepackage{colortbl}

\usepackage{txfonts}
\usepackage[utf8]{inputenc}

\usepackage[colorlinks=true, pdfstartview=FitV, linkcolor=green, citecolor=green, urlcolor=green, bookmarksopen]{hyperref}

\makeatother

\begin{document}

\preprint{$\today$}

\title{Field-driven Reversal Models in Artificial Spin Ice}

\author{Gary W. Paterson}
    \email{Dr.Gary.Paterson@gmail.com}
    \affiliation{SUPA, School of Physics and Astronomy, University of Glasgow, Glasgow G12 8QQ, UK.}
    \affiliation{James Watt School of Engineering, Electronics \& Nanoscale Engineering Division, University of Glasgow, Glasgow, G12 8QQ, UK.}
\author{Gavin M. Macauley}
    \altaffiliation[Current addresses: ]{Laboratory for Mesoscopic Systems, Department of Materials, ETH Zurich, 8093 Zurich, Switzerland; Laboratory for Multiscale Materials Experiments, Paul Scherrer Institute, 5232 Villigen PSI, Switzerland.}
    \affiliation{SUPA, School of Physics and Astronomy, University of Glasgow, Glasgow G12 8QQ, UK.}
\author{Rair Mac\^edo}
    \affiliation{James Watt School of Engineering, Electronics \& Nanoscale Engineering Division, University of Glasgow, Glasgow, G12 8QQ, UK.}

\date{\today}

\begin{abstract}
We investigate a set of topological arrangements of individual ferromagnetic islands in ideal and disordered artificial spin ice (ASI) arrays in order to evaluate how aspects of their field-driven reversal are affected by the model used.
The set contains the pinwheel and square ice tilings, and thus a range of magnetic ordering and reversal properties are tested. 
We find that a simple point dipole model performs relatively well for square ice, but it does not replicate the properties observed in recent experiments with pinwheel ice.
Parameterization of the reversal barrier in a Stoner-Wohlfarth model improves upon this, but fails to capture aspects of the physics of ferromagnetic coupling observed in pinwheel structures which have been attributed to the non-Ising nature of the islands.
In particular, spin canting is found to be important in pinwheel arrays, but not in square ones, due to their different symmetries.
Our findings will improve the modelling of ASI structures for fundamental research and in applications which are reliant upon the ability to obtain and switch through known states using an externally applied field.
\end{abstract}

\keywords{artificial spin ice, pinwheel, square, field-driven, avalanche, disorder}

\maketitle

\section{Introduction}
The plethora of ways in which interacting single domain nanomagnetic islands can be arranged into artificial spin ice (ASI) arrays has led to great interest in them both for research and potential applications in a number of areas, including tunable radio frequency magnonic metamaterials,~\cite{Lendinez_2019_JPCM_dynamics_rev, Gliga_2020_ASI_dynamics_rev, Skjaervo_2020_NatRevs_asi, Kaffash_PLA_2021_nanomagnonics} and for data storage and neuromorphic computation.~\cite{Imre_science_2006_ASI_computation, Jensen18_asi_computation, Chern_pra_17_hall_circuits, Arava_PRA_2019_asi_relax, Caravelli_arxiv_2018_bool_gates, Arava_2018_sq_logic, Hon2021_APE_Kagome_reservoir}
Fundamental to almost all applications of ASIs is the ability to reconfigure the magnetization of the system through application of heat or an external field.
The ability to accurately model the response of ASIs to such stimuli is often key to understanding the origin of their functionality and to extracting information from experimental data.
This ability can also be critical to designing new configurations, in testing and developing new protocols before performing them in experiments, and in modelling procedures that would be difficult or impractical to realise, such as studies of criticality.~\cite{Mengotti_NatPhys_2011, Chern2014_njp_sq_kg_avalanches}

The assumed single-domain nature of each island in ASIs was one of the original attractions~\cite{Wang_nature_2006_square_ASI} of ASIs as analogues of Ising~\cite{Ising_1925em} systems.
That such systems are often amenable to being modelled using the point dipole approximation has made this type of calculation a mainstay of ASI studies for its simplicity and computational efficiency.
In this approximation, the net moment of each island is represented by a single point dipole, which is constrained to lie either parallel or anti-parallel to the island long axis.
It is remarkable that such a simple model, or ones based on dumbbells or multipole expansions,~\cite{Moller2009_PRB_multipole, Di_Pietro_Mart_nez_2020_disorder} can capture the essence of the complex physics of many systems.

However, there is an increasing appreciation that the non-Ising nature of macrospins can have an influence on the collective behavior of ASIs, particularly in field-driven cases, or when the dipolar interaction between all neighboring islands is comparable to each other or weak.
These include the effects of magnetostatic bias on island reversal processes,~\cite{Porro_JAP_2012_vortex_islands} domain wall propagation in connected systems,~\cite{Ladak_2012_njp_honeycomb} charge propagation in isolated systems,~\cite{Rougemaille_NJP_2013_chiral_end_states} end-state signatures in high-frequency dynamics,~\cite{Gliga_PRB_2015_sq_vertex_endstate, PANAGIOTOPOULOS_jmmm_2017_kagome_RF} modification of coupling in pseudo-one-,~\cite{Nguyen_PRB_2017_1d_phase_diag_end_states} two-,~\cite{Velo2020_kagome_micromag, Schanilec2020_PRL_Kagome_freezing} and three-dimensional~\cite{Perrin2016_nature_square} systems, and the importance of edge roughness.~\cite{Kohli2011_PRB_edge_roughness_SQ, Velo2020_kagome_micromag}
While these non-Ising properties offer a computational challenge, they also provide an additional means by which to tailor the properties of the system.
For example, it has very recently been shown that a small change to the shape of a kagome lattice allows the elusive ground state to be reached~\cite{Schanilec2020_PRL_Kagome_freezing}. 

One system where non-Ising interactions have been shown to be the route to novel phenomena observed during field-driven reversal is the pinwheel geometry.
This system is formed by rotating each island in square ASI~\cite{Wang_nature_2006_square_ASI} by some angle.
While the thermal ground state of square ice is well-known to exhibit antiferromagnetic (AFM) ordering,~\cite{morgan_NatPhys_2010_sq_AFM_GS} ferromagnetic (FM) ordering is preferred when the islands in square ice are rotated by $\pi/4$.~\cite{Macedo_PRB_2018}
Examples of a pinwheel and square ice repeat unit are shown in \textbf{Figure~\ref{fig:geom_single_astroid}(a)} and \textbf{\ref{fig:geom_single_astroid}(b)}, respectively, where we also define the rotation angle, $\alpha$, as 0$\mathrm{^\circ}$ for pinwheel.
The ability to alter the mesoscale magnetic ordering, the magnetic texture, and the dimensionality of the reversal mechanism from one-dimensional Dirac strings in square~\cite{Morgan_njop_2011_sq_reversal} to two-dimensional in pinwheel~\cite{Li_acs_nano_2019} by varying $\alpha$ makes this system particularly interesting for applications and studying models of phase transitions.~\cite{Macauley_PRB_thermalised_asi, Massouras_PRB_2020_pw_fm_afm}
The reconfigurability of this spin system has been used to create modulating fields in hybrid devices,~\cite{Gliga_2020_PRA_ASI_dynamics, Lyu2020_nanoLett_SC_PW} while the complex field-driven spatio-temporal patterns it supports has been identified as particularly interesting for reservoir computing due to the presence of non-linearity and memory in the system.~\cite{Jensen18_asi_computation, Jensen2020_PW_reservoir}

Importantly, the reversal process in pinwheel ASI~\cite{Li_acs_nano_2019} is not purely due to changes in the distribution of dipolar coupling strengths,~\cite{Macedo_PRB_2018} but it has also been attributed to a breaking of the Ising nature of the island magnetization.~\cite{Paterson_2019_PRB_pinwheel_FM}
In particular, both the bending of island magnetization in `end-states' that form at the point of reversal in field-driven processes, and the incorporation of an angle dependent energy barrier significantly influence the collective properties.
Along with the finite island size, the non-Ising aspects of the island properties in the pinwheel arrangement creates a strong coupling between nearest neighbor islands that gives rise to pseudo-exchange effect and emergent anisotropies which are offset from the geometrical axes of the array.
Such magnetostatic bias effects have been seen on other systems,~\cite{Porro_JAP_2012_vortex_islands, Nguyen_PRB_2017_1d_phase_diag_end_states, PANAGIOTOPOULOS_jmmm_2017_kagome_RF, PANAGIOTOPOULOS_PB_2016_kagome_bias} and the complex collective interactions in pinwheel geometries have been attributed to chiral effects.~\cite{Gliga_nmat_2017, Wyss_2019_acsnano_chiral}

The range of properties exhibited by the square--pinwheel continuum of spin ice geometries makes it an ideal test bed to study different models of nanomagnetic systems, which is the focus of this work.
We investigate the use of different dipolar models using analytical and numerical~\cite{Jensen_2020_flatspin} calculations for simulating the field-driven reversal paths of square and pinwheel arrays, and compare the results to those from the micromagnetic MuMax3 package.~\cite{mumax3_2014, Leliaert_2018_mumax3_advances}
Whether the additional physics of micromagnetic models are significant or not depends on the system and properties being investigated.
The different models used in this work are defined in Section~\ref{sec:models}.
In Section~\ref{sec:anisotropy} we explore how the inter-island coupling differs in the models used, and how anisotropy can arise from collective behavior in the pinwheel system only when a field-angle dependent barrier is employed.
The importance of spin canting -- in essence, a simplified model for the effect of end-states -- is investigated in Section~\ref{sec:spin_canting}.
There, we show that it modulates the coercive fields of all arrays, but that it only affects the anisotropy axis of pinwheel ones.
In Section~\ref{sec:ideal_revsal_paths} we examine the reversal paths of ideal arrays, while in Section~\ref{sec:disorder} we investigate the influence of disorder and find that the stronger inter-island interactions reproduced in the micromagnetic models impart a significant degree of robustness against its influence.

Our findings not only shed light on aspects of the square--pinwheel system, but also highlight important features that are common to almost all ASI systems: that the use of an appropriate angle dependent reversal barrier can have a significant effect on the coupling in field-driven processes; and that spin-canting and end-states provide additional degrees of freedom that can influence the emergent properties.
We expect that the inclusion of these properties may improve the modelling of a range of ASI structures.

\section{Interaction Models}
\label{sec:models}
\begin{figure}[hbt]
  \centering
      \includegraphics[width=8.5cm]{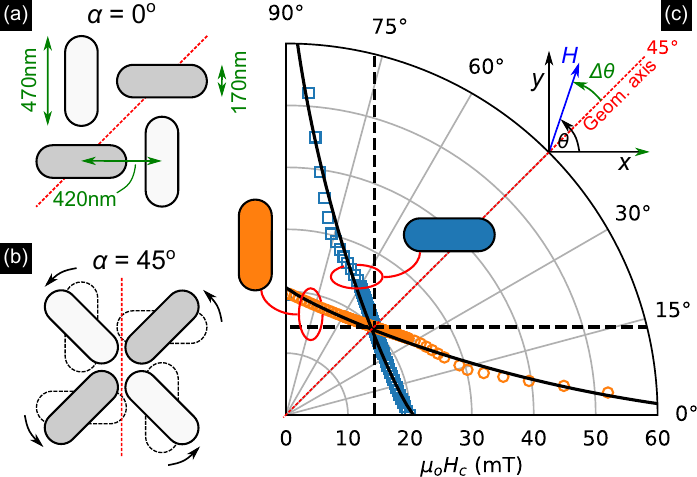}
      \caption{Arrangement of macrospins in (a) pinwheel and (b) square tilings of four islands, showing the island and array sizes, and the island rotation angle, $\alpha$.
      The grey / white islands represent subarrays wherein the islands are parallel to one another.
      (c) Coercive field, $H_c$, astroids as a function of the applied field angle, $\theta$, for isolated horizontal and vertical islands, for the micromagnetic (MM) model (symbols), point-dipole (PD) model (dashed lines), and fits of the Stoner-Wohlfarth point-dipole (SW-PD) model (solid lines) to the MM data.
      The inset to (c) defines the applied field direction with respect to the $x$-axis for the islands shown, $\theta$; the geometrical axis at $\theta = 45\mathrm{^\circ}$; and the direction of offsets from the geometrical axis, $\Delta \theta$.
      In all panels, the red dashed lines show the geometrical axis.
      The MM data was taken from \cite{Paterson_2019_PRB_pinwheel_FM_data}.}
      \label{fig:geom_single_astroid}
\end{figure}

In this section, we define the different models and ASI tilings used throughout this work.
The pinwheel geometry for a small array is shown in Figure~\ref{fig:geom_single_astroid}(a).
As depicted in the figure, the islands were taken to be stadium shaped.
The degree of Ising breaking, and thus the results dependent upon it, will vary with island size.
Here, we use islands of length 470~nm, width 170~nm, and thickness 10~nm, arranged in a lattice with a nearest neighbor distance of 420~nm, in order to match our earlier work on this system.~\cite{Li_acs_nano_2019, Paterson_2019_PRB_pinwheel_FM}
A set of geometries is defined by island rotation angle, $\alpha$, from pinwheel at $\alpha = 0^{\mathrm{o}}$ to square at $\alpha = 45^{\mathrm{o}}$, as shown in Figure~\ref{fig:geom_single_astroid}(b).
The models are, in part, defined by the island reversal barrier, which we quantify by the angular dependence of the island's coercive field.
We consider three main models of the two-dimensional (2-D) arrays: two point dipole models with different island reversal criteria, and a micromagnetic (MM) one.
In the MM model, interactions between real spins are accounted for by discretisation of the magnetisation in the continuum limit with a cell size around or less than the exchange length, whereas in the point dipole models, each entire ferromagnetic island is replaced by a single macrospin with fixed properties, and thus each model has different prospects for capturing complex interactions and emergent phenomena.

In the point dipole models, each real 2-D island is represented by a single point dipole with a moment equal to that of the saturated extended island, and which is constrained to lie along the long-axis of the real island.
We will refer to the first of these models as \emph{the} point dipole (PD) model, and adopt the reversal criterion of the net field along the long-axis of the island exceeding a threshold, as commonly used in field-driven square and kagome arrays.~\cite{budrikis_PRL_2012_sq_disorder, Chern2014_njp_sq_kg_avalanches}
The second reversal criterion we consider is a parametrization of the reversal barrier astroids~\cite{Thiaville_PRB_2000_astroids} in a Stoner-Wohlfarth (SW) model of coherent rotation within an extended volume,~\cite{Stoner_Wohlfarth_1947} following that recently implemented in the Python package flatspin;~\cite{Jensen_2020_flatspin} we will refer to this as the Stoner-Wohlfarth point-dipole (SW-PD) model.
In this model, it is important to note that the spins themselves are fixed and not allowed to rotate (other than an instantaneous 180$^{\mathrm{o}}$ reversal).
For the micromagnetic calculations, we use the MuMax3 finite-difference simulation program~\cite{mumax3_2014, Leliaert_2018_mumax3_advances} which allows for more realistic description of the islands in the modelling by including their true shape and by allowing the magnetization of each island to adapt to the field distribution across its volume in accordance with the shape anisotropy of the island and the exchange strength of the material.

In a previous work, we found that the experimentally observed reversal of pinwheel arrays~\cite{Li_acs_nano_2019} could not be replicated using a simple PD model, but that the main features could be reproduced through MM modelling.~\cite{Paterson_2019_PRB_pinwheel_FM}
It was recently reported that the use of the SW-PD model also allows the magnetization reversal of pinwheel ASI to be reproduced.~\cite{Jensen_2020_flatspin}
While the SW-PD model is a substantial improvement on the PD model, we will show that not all magnetization properties are reproduced in this model.
In exploring why this is the case, we also provide suggestions for how aspects of the model can be improved without resorting to micromagnetics.

The coercive field, $H_c$, astroids from MM simulations for isolated islands of each sub-lattice are shown as symbols in Figure~\ref{fig:geom_single_astroid}(c).
The magnetic parameters of the islands were based on the material properties of permalloy, namely an exchange stiffness of 13 pJ$\,$m$^{-1}$, a saturation magnetization of 800 kA$\,$m$^{-1}$, and a Gilbert damping parameter of 0.02.
The applied field direction is defined by the angle $\theta$ from the $x$-axis, as shown in the top right inset.
The dashed red line shows the geometrical axis, where the field would be applied at equal angles to the respective symmetry axes of the islands on both sub-lattices; these sub-lattices are differentiated by their light and dark grey color in Figure~\ref{fig:geom_single_astroid}(a) and \ref{fig:geom_single_astroid}(b).

The dashed black lines in Figure~\ref{fig:geom_single_astroid}(c) show the PD criterion for switching of $-\bm{M} \cdot \bm{H} > H_c$, where $\bm{M}$ is the net moment of the macrospin and $\bm{H}$, the net field at the spin location, is composed of the external field and dipolar contributions from the other islands.
The value of 14.2~mT was chosen to match the MM value of 20.1~mT at an applied field angle of 45$^{\mathrm{o}}$.
This figure makes clear one the the key differences between the PD and other models: the PD barrier is only a good approximation of the MM one when the fields are applied close to the geometrical axis of the array; indeed, the PD barrier goes to infinity at high angles.
However, will show that the effect of coupling between islands is significantly different for the two models, even when the field is applied close to the geometrical axis.

The form of the SW equation proposed in \cite{Jensen_2020_flatspin} is overdetermined for our data, so we instead use the following parametrization to describe the astroids
\begin{equation}
    \left( \frac{H_{\parallel}}{b H_k} \right)^{\frac{2}{\gamma}} + \left( \frac{H_{\perp}}{H_k} \right)^{\frac{2}{\beta}} = 1,
    \label{eqn:sw}
\end{equation}
where $H_{\parallel}$ and $H_{\perp}$ are the field components parallel and perpendicular to the island long axes, $H_k$ represents the field required to overcome the reversal barrier formed by the short axis of the island, and $b$, $\gamma$ and $\beta$ are fitting parameters whose values were determined to be 0.251$\pm$0.004, 3.39$\pm$0.14 and 2.27$\pm$0.11, respectively.
In principle, the value of $H_k$ could be found by fitting, but to reduce the degrees of freedom so that errors could be estimated, it was fixed at the value suggested by our micromagetic calculations of 82~mT.
The fitted astroids are shown by the solid lines in Figure~\ref{fig:geom_single_astroid}(c), and are a good approximation for the MM data for these isolated islands.
We note here that island reversal in the MM model is not by pure coherent rotation but, nevertheless, the angle dependence of the barrier is well approximated by the SW equation.

Having defined these models, we now apply each of them separately to track important properties of coupled arrays.
In particular, we will consider the inter-island coupling and any emergent anisotropies, the influence of spin canting, the reversal paths for ideal arrays, and how the results predicted by each model are affected by disorder.

\section{Inter-island Coupling and Emergent Anisotropy}
\label{sec:anisotropy}
In ASI tilings, the dipolar fields from other islands comprise effective fields that may give rise to anisotropies in the collective behavior.
Whether global array anisotropies exist depends on anisotropy of the individual islands and the symmetry of the array formed by them.
In the pinwheel geometry, there exists an anisotropy that is misaligned with the geometrical axis, in part due to end-states of the constituent islands, and this changes depending on the array edge termination, whereas, for the square geometry, end-states give rise to a degree of frustration, but the arrays do not show any misalignment of the anisotropy axis.~\cite{Paterson_2019_PRB_pinwheel_FM}
As a consequence, we focus here on the pinwheel structure.
We begin by looking at the unit formed from four islands, for which we need only consider one island from each sub-lattice due to the symmetry of the system (larger arrays are considered later).

\begin{figure}[hbt]
  \centering
      \includegraphics[width=8.5cm]{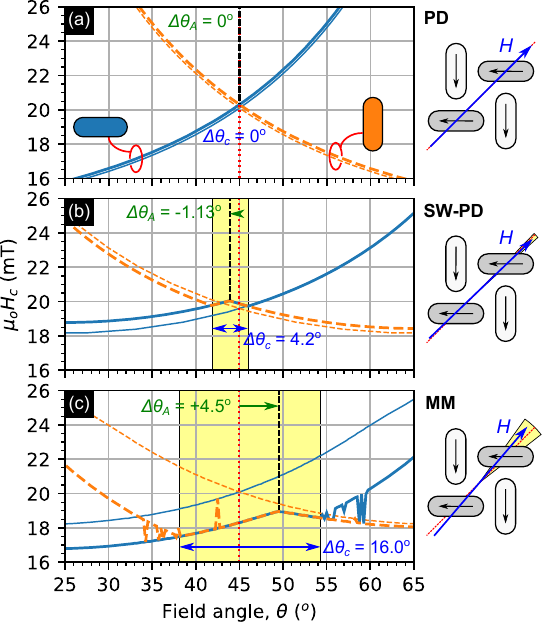}
      \caption{Astroids for the leftmost islands in a pinwheel unit (thick lines) and isolated islands (thin lines) for the (a) PD, (b) SW-PD, and (c) MM models.
      The parameters $\Delta \theta_A$ and $\Delta \theta_c$ annotated in each panel are the anisotropy axis misalignment angle and the angular range of the strong coupling regime (also shown by the yellow shaded regions), respectively.
      The schematics on the right display the equivalent configurations for the three datasets, with the field applied at the anisotropy axes.
      The PD and SW-PD data for coupled arrays were calculated in the flatspin package, whereas those for isolated islands were calculated analytically.
      The MM data was taken from \cite{Paterson_2019_PRB_pinwheel_FM_data}.
      The noise in this data arose from occasional metastable states that went uncorrected and do not otherwise affect the results.}
      \label{fig:pw_unit_astroid}
\end{figure}

\textbf{Figure~\ref{fig:pw_unit_astroid}} shows two sets of astroids for two islands of a pinwheel unit for the three different models considered, one set without any coupling (thin lines) and one set with coupling (thick lines).
In all models, when the islands are uncoupled, the astroids cross at applied field angles of 45$^{\mathrm{o}}$, as expected, and at very similar $H_c$ values.
For the MM model [Figure~\ref{fig:pw_unit_astroid}(c)], coupling between islands causes three effects: $H_c$ reduces in magnitude; the astroids snap together over a range of angles, $\Delta \theta_c$, of 16.0$^{\mathrm{o}}$ where the coupling is strong (these regions are marked by a yellow background); and the field angle at which the astroids intersect and where $H_c$ is maximal is misaligned from the geometrical axis.
This offset in angle, $\Delta \theta_A = +4.5^{\mathrm{o}}$, defines the anisotropy axis of the array.
This property in larger arrays is discussed later with reference to \textbf{Figure~\ref{fig:sw_sqo_pw_h_nets}}.
For all models, the schematics on the right of the figure display the configuration with the field applied at the anisotropy axis.

For the PD and SW-PD models, the dipolar field magnitude and angle from each macrospin is unaffected by the field applied to it.
This is one of the greatest weaknesses of the models, but it also allows us to easily examine them analytically.
For the PD model, the biasing of the energy barrier (in Joules) of each island in the four island unit, $-\bm{M} \cdot \bm{H}$, from the dipolar field from all other islands, $H_{dip}$, is identical:
\begin{equation}
    \Delta E(\alpha) = \frac{MD}{4\sqrt{2}} \left(1 + 3\sin(2\alpha)\right), \label{eqn:pdp_E}
\end{equation}
where $D = \mu_0 M / (4 \pi a^3)$ is the dipolar field coupling constant in which $\mu_0$ is the permeability of free space, $M$ is the net island moment magnitude (A$\,$m$^2$), and $a$ is the first nearest neighbor distance (for our arrays: $a$ = 420~nm, $D$ = 7.958$\times 10^{-4}$~T).
Since Eq.~(\ref{eqn:pdp_E}) applies to all islands, no misalignment of the anisotropy axes with respect to the geometrical axes can exist in the PD model for \emph{any} island rotation angle.
The PD astroids of the \emph{coupled} islands, calculated using the flatspin package [Figure~\ref{fig:pw_unit_astroid}(a)] confirm the result expected from Eq.~(\ref{eqn:pdp_E}): there is no anisotropy axis misalignment, no strong coupling regime, and there is a small increase in $H_c$.
In all flatspin simulations, the other parameters were: $\mathtt{\alpha_{flatspin}}$ = 7.958$\times 10^{-4}$~T,  \texttt{lattice\_spacing}=$\sqrt{2}$, and \texttt{neighbor\_distance} = 10.
Comparing the PD curves [Figure~\ref{fig:pw_unit_astroid}(a)] to those from MM [Figure~\ref{fig:pw_unit_astroid}(c)] also makes clear just how much more rapidly they increase with angle than they should do due to the simplicity of the reversal criterion, as discussed above.

For the SW-PD model, no simple analytical forms exist for the anisotropy axis, so we solve the system of equations numerically using the Python CAS system SymPy.~\cite{sympy_2017}
The presence of dipolar fields effectively shifts the SW astroids, thus imparting an offset in the anisotropy axis and a change in the coercive field value.
For our sample, $\Delta \theta_A$ is -1.122$^{\mathrm{o}}$ and $\mu_o H_c$ increases a small amount (from 19.587~mT to 20.054~mT).
The flatspin calculated astroids for the SW-PD model [Figure~\ref{fig:pw_unit_astroid}(b)], reproduce the anisotropy axis misalignment and the change in coercive fields compared to uncoupled islands.
However, these values are both of the wrong sign and of much smaller magnitude than the MM values.
The plots also show that the strong coupling regime is much narrower than in the MM model, with $\Delta \theta_c$ being only 4.2$^{\mathrm{o}}$ wide in the SW-PD model.


While the SW-PD model is a great improvement upon the PD one, significant differences compared to the MM one remain.
This is especially true for field-driven cases where the moment within each macrospin can rotate in response to the external field.
Next we will show in the SW-PD model applied to the pinwheel geometry that it is precisely this effect of `spin canting' that is directly responsible for the sign of the anisotropy axis misalignment and that this also contributes to determining its magnitude.

\section{Spin Canting}
\label{sec:spin_canting}
\begin{figure}[hbt]
  \centering
      \includegraphics[width=8.0cm]{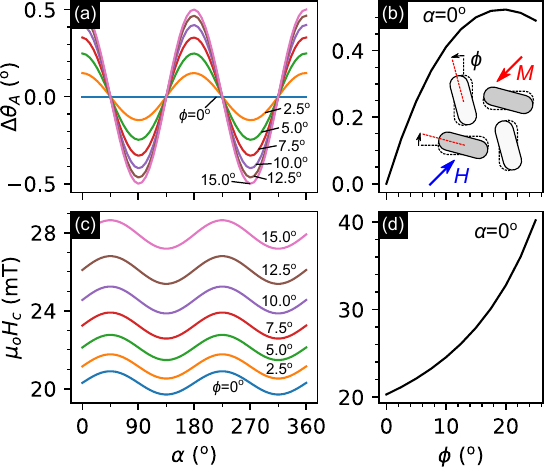}
      \caption{The effect of spin canting on (a) and (b) the anisotropy axis misalignment and (c) and (d) the coercive field in the PD model.
      (a) and (c) show these properties as a function of island rotation, $\alpha$, at selected canting angles, $\phi$, while (b) and (d) show the same properties for the pinwheel geometry ($\alpha = 0\mathrm{^\circ}$).
      The schematic inset to (b) defines the canting angle, with the point-dipoles drawn as 2-D islands, and shows the applied field, $H$, and the net magnetisation, $M$. }
      \label{fig:pd_phi_alpha}
\end{figure}

Subject to an external field applied along the geometrical axis of an ASI array, the net moment of each ferromagnetic island slightly rotates within the fixed island to minimise the Zeeman contribution to the net energy, with the rotation occurring in opposite directions for each sub-lattice.~\cite{foot1}
This spin canting has an important effect on the interactions between the macrospins and the emergent properties of pinwheel arrays.~\cite{Paterson_2019_PRB_pinwheel_FM}
The possibility of uniform spin canting in response to an applied field is included in MM models, but is absent in the PD and SW-PD models.
In the following, we explore this degree of freedom in the point-dipole models by introducing into our analytical calculations the parameter $\phi$, describing the canting angle of the point-dipoles.
This situation is depicted in the schematic inset to \textbf{Figure~\ref{fig:pd_phi_alpha}(b)}, where the point-dipoles are drawn as 2-D islands.
Because the moment of islands in each sub-lattice rotate in opposite directions, the angle of the net magnetisation remains unchanged.
The degree of canting will depend on the external field and could be estimated from MM simulations. 
For simplicity, we examine the case of equal spin canting, which approximates the case where the field is applied along a geometrical axis.

For the PD model, the inclusion of canting breaks the degeneracy of the reversal barrier in Eq.~(\ref{eqn:pdp_E}).
The biasing of the energy barrier from the dipolar field from all other islands now takes the form:
\begin{subequations}
\begin{equation}
    \Delta E_{TL}(\alpha, \phi) = \frac{MD}{4\sqrt{2}} \left( 1 + 3\sin(2\alpha + 2\phi) - 4\sqrt{2} \sin(2\phi) \right) \label{eqn:pdp_phi_E0}\\
\end{equation}
for the top left island, and
\begin{equation}
    \Delta E_{BL}(\alpha, \phi) = \frac{MD}{4\sqrt{2}} \left( 1 + 3\sin(2\alpha - 2\phi) - 4\sqrt{2} \sin(2\phi) \right) \label{eqn:pdp_phi_E2}\\
\end{equation}
\end{subequations}
for the bottom left one.
For there to be no anisotropy axis misalignment in the PD model, $\Delta E_{TL}(\alpha, \phi) = \Delta E_{BL}(\alpha, \phi)$, and this occurs for two conditions.
The trivial solution is the one we saw in the previous section: $\phi = 0$ for any $\alpha$ value.
The second solution is $\alpha = \pi/4 + n\pi/2$, where $n \in \mathbb{Z}$, for any $\phi$ value.
In other words, square units can never have a misaligned anisotropy axis, even if the spins are canted.
For all other $\alpha$ values, $\Delta \theta_A$ is non-zero if there is spin canting.

To examine the properties of the models further, we must once again turn to numerical methods.
The results for the PD model, using the parameters for our system, are shown in Figure~\ref{fig:pd_phi_alpha}.
As $\phi$ increases from zero, the magnitude of $\Delta \theta_A$ increases from zero whilst maintaining a sinusoidal dependence on $\alpha$ [Figure~\ref{fig:pd_phi_alpha}(a)], as a result of the handedness of the pinwheel unit periodically reversing.
The misalignment amplitude peaks for the pinwheel unit, and this dependence on $\phi$ is shown in Figure~\ref{fig:pd_phi_alpha}(b).
The maximum value of $\Delta \theta_A$ occurs at a $\phi$ value of $\sim$20$\mathrm{^\circ}$ and is 0.52$\mathrm{^\circ}$.
Spin canting also causes the coercive field to increase at all spin rotations [Figure~\ref{fig:pd_phi_alpha}(c)].
Figure~\ref{fig:pd_phi_alpha}(d) shows this dependence on $\phi$ for the pinwheel geometry, where it can be seen that the coercive field increases extremely rapidly.
Including spin canting in the PD model thus further increases the deviation of important properties from the MM results. 
A contributory reason for this is that a significant fraction of the dipolar field is simply ignored in the PD model, and we therefore might expect the SW-PD model to produce better results upon the inclusion of spin canting.

\begin{figure}[hbt]
  \centering
      \includegraphics[width=8.0cm]{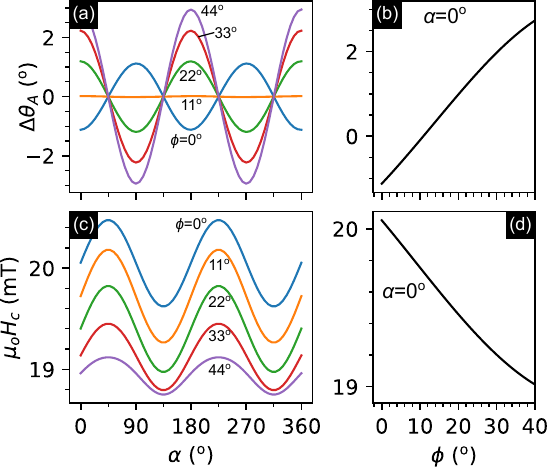}
      \caption{The effect of spin canting on (a) and (b) the anisotropy axis misalignment and (c) and (d) the coercive field in the SW-PD model. (a) and (c) show these properties as a function of island rotation, $\alpha$, at selected canting angles, $\phi$, while (b) and (d) show the same properties for the pinwheel geometry ($\alpha = 0\mathrm{^\circ}$).}
      \label{fig:sw_phi_alpha}
\end{figure}

The equivalent properties of the PD model of Figure~\ref{fig:pd_phi_alpha} are shown in \textbf{Figure~\ref{fig:sw_phi_alpha}} for the SW-PD model.
Spin canting in this model improves the anisotropy axis misalignment, which, for the pinwheel array, increases from negative values to positive ones with increasing $\phi$ [Figure~\ref{fig:sw_phi_alpha}(a) and \ref{fig:sw_phi_alpha}(b)].
For our system, a canting angle of around 11$\mathrm{^\circ}$ almost completely removes the anisotropy angle misalignment for all $\alpha$ [Figure~\ref{fig:sw_phi_alpha}(a)], but the MM value of +4.5$\mathrm{^\circ}$ cannot be reached at realistic canting angles.
The precise effective canting angle that will be reached before a real ferromagnetic island reverses depends on many factors, including the geometry of the islands and the exchange length of the material, `imperfections' such as the edge roughness and material granularity, and the interactions between islands.
Unlike in the PD model, spin canting in the SW-PD model reduces the coercive field [Figure~\ref{fig:sw_phi_alpha}(c) and \ref{fig:sw_phi_alpha}(d)], but less than needed to match the MM results shown in Figure~\ref{fig:pw_unit_astroid}(c) of $\sim$1.5~mT at angles around the geometrical axis.

Anisotropy axis misalignment in the four island square array is also forbidden in the SW-PD model.
The reason for this is that both the net dipolar field magnitude and the magnitude of its angular offset from the symmetry axis of the macrospin remains balanced across the macropsins for all values of $\phi$.
This result is expected from symmetry considerations, and is the main reason why the PD and SW-PD models can serve as a reasonable approximation for the interactions in field-driven square ASI arrays.
The advantage of the SW-PD model over the PD one for square arrays is that it incorporates a more accurate coercive field $\theta$ dependence, which will also be an important consideration for simulations including disorder.

While spin canting does not include all of complex internal magnetization properties of each island, the absence of it in the SW-PD model is one of the main deficits of the model, which affects the collective behavior, including the anisotropy axis and the coercive field.
Much like how the SW-PD model improves upon the switching criterion by use of a parameterization of the SW astroid, the complex field magnitude and angle dependence of each island may be determined from micromagnetic simulations and parameterised for inclusion in optimised numerical calculations.
During a simulation, the degree of spin canting could be determined by the net field on each island iteratively or, since the barrier to reversal is generally much greater than the dipolar field magnitudes, the spin canting may simply be approximated using the external field alone.
This additional degree of freedom may also improve the modelling aspects of the square system, where end-states impart a degree of disorder [see Section~\ref{sec:ideal_revsal_paths}].

We note that the spin canting findings in this section are not strictly valid for larger arrays due to there existing dipolar field contributions from different configurations of neighbors lying at different angles to each island.
However, for all arrays considered, the symmetries are unmodified, and thus the relative anisotropy axis alignment or misalignments should be present in larger arrays, albeit at different magnitudes (an example of this is given in Figure~\ref{fig:sw_sqo_pw_h_nets}, discussed later).
The additional degree of freedom of spin-canting in an applied field acts somewhat similar to a spring under a load, and so we would expect it to increase the angular width of the strong coupling regime and that it may also influence aspects of domain propagation across arrays.~\cite{foot2}

Some of the reduced coupling seen in the modelling done here will also be due to condensing the magnetization in to a single point, as observed in other systems~\cite{Rougemaille_PRL_2011_kag}.
While artificially increasing the moment of each island will not alter the first NN bond energy, $J$, it will increase the coupling during reversal due to the increased local dipolar field, at the cost of modifying the coupling and bond energy with further distant neighbors due to the different distribution of fields.
A similar effect is likely to be seen by artificially reducing the lattice spacing.
These aspects of the interactions may be improved by adopting a dumbbell model where the finite length of the islands is incorporated~\cite{Moller2009_PRB_multipole, Di_Pietro_Mart_nez_2020_disorder}, or each island may be broken into several point dipoles, as others have done in a different system~\cite{Yu_2017_AIP_Adv_mc}.
Spin canting can be included in all three models, by rotating the point dipole, the dumbbell, or the end spins, and the combination of the effects is likely to further improve modeling of the coupling and all the important properties reliant upon it.
These potential improvements could be added to an implementation of the SW-PD model to improve the accuracy of the calculations whilst still maintaining much of the speed advantages that point dipole calculations have over the more advanced MM models.
Alternatively, the use lookup tables of relevant properties, either derived directly from micromagnetics or altered to produce suitably modified interactions in whatever version of the point-dipole model is used, is likely to prove beneficial, at the cost of a degree of complexity.
As none of these features are present in currently available packages, we will continue with our model evaluation using the fixed single point-dipole models.

\section{Ideal Reversal Paths}
\label{sec:ideal_revsal_paths}
For many of the potential applications of the pinwheel and other arrays, the precise reversal path is important, so it is critical to know that the simulations accurately capture this aspect of the collective behavior.
Our previous MM simulations of pinwheel arrays showed that the anisotropy axis misalignment and coercive field changed with increasing array sizes up to arrays of 32 islands, formed from a 4$\times$4 sub-lattice interleaved with one of the same size~\cite{Paterson_2019_PRB_pinwheel_FM}.
This marks the point where the islands at the corners of the arrays effectively have a full set of neighbors.

\begin{figure}[hbt]
  \centering
      \includegraphics[width=8.0cm]{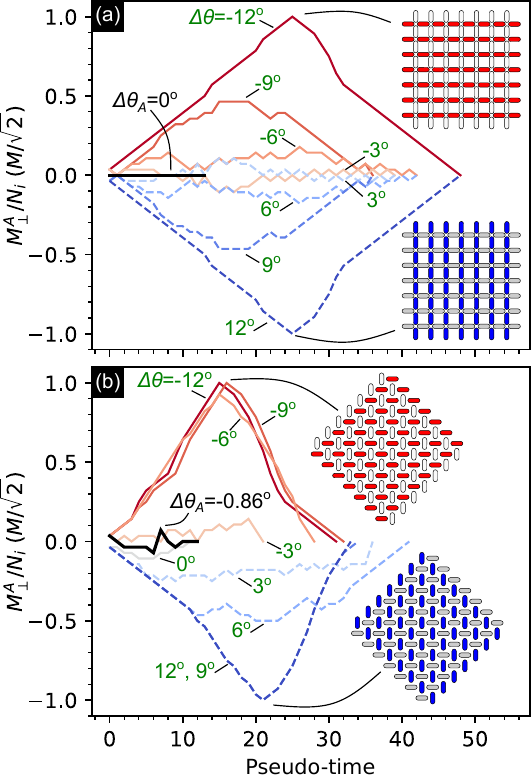}
      \caption{Net array moment, $M^A$, perpendicular to the geometrical axis during a SW-PD modelled reversal of (a) an open square and (b) a pinwheel structure as a function of pseudo-time, at different applied field directions, $\Delta \theta$, with respect to the geometrical axis.
      We define pseudo-time as one iteration of the reversal of island(s) and an updated dipolar field calculation.
      Each array is formed by an 8$\times$7 sub-lattice interleaved with a 7$\times$8 one.
      The colored islands in the insets in each panel show the islands that have reversed at each extrema of magnitude $\pm$1.
      The thick black lines show the reversal path for the case of the external field applied along anisotropy axis, $\Delta \theta_A.$
      In both panels, the geometrical axis is directed north-east and the field step was 1~$\mu$T.
      $N_i$ is the number of islands, and $M$ is the moment of a single island.}
      \label{fig:sw_sqo_pw_h_nets}
\end{figure}

To evaluate the reversal paths here, we use arrays formed by a 8$\times$7 sub-lattice interleaved with a 7$\times$8 one to form an overall array with a rotationally symmetric edge.
Examples of the open edged square and the pinwheel arrays of this size are shown in the insets to Figure~\ref{fig:sw_sqo_pw_h_nets}(a) and \ref{fig:sw_sqo_pw_h_nets}(b), respectively.
These 112 island arrays have roughly four times the number of islands of the one at which the plateau is reached in the pinwheel geometry, and so will incorporate some properties of the bulk.

\begin{figure}[!ht]
  \centering
      \includegraphics[width=8.5cm]{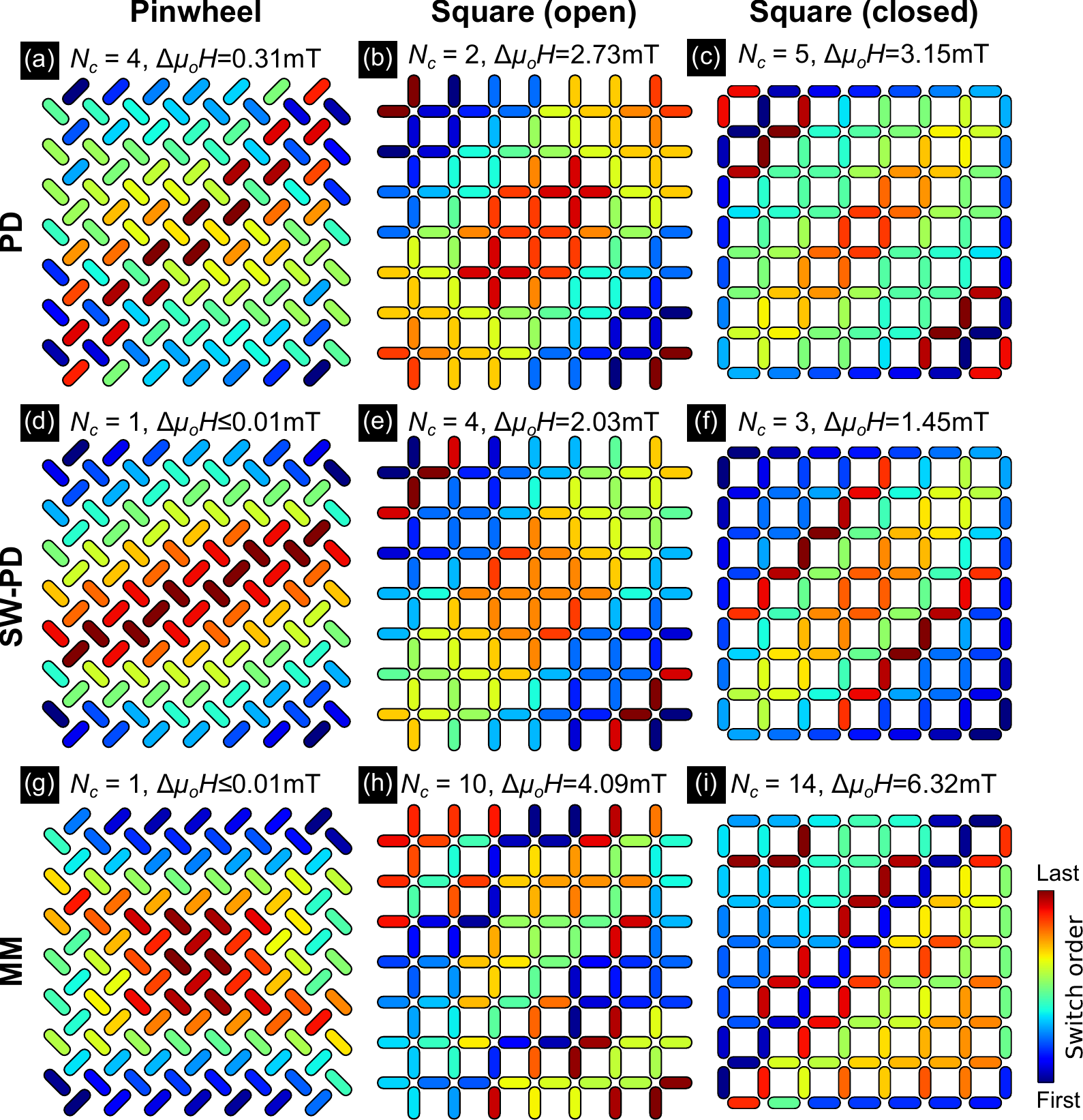}
      \caption{The reversal paths of ideal pinwheel and square arrays (in columns) simulated using the PD (row 1), SW-PD (row 2) and MM (row 3) models with a field step of 10~$\mu$T.
      The annotations show the number of cascades, $N_c$, and the field range $\Delta \mu_o H$ of the reversal.
      The field was applied along the respective anisotropy axis for each array (exactly north-eastwards for the square arrays and approximately eastwards for the pinwheel arrays).
      For the MM simulations of the square arrays, the magnetization was allowed to evolve in time from a field of 0~T, so that the end-states formed naturally.
      Videos showing the reversal orders are included in Supporting Information.}
      \label{fig:ideal_reversal_model_maps}
\end{figure}

For the MM simulations here, we modify our simulation methodology reported before~\cite{Paterson_2019_PRB_pinwheel_FM} in order to allow the internal interactions of the system to have the maximum effect.
We do this by stepping the field by the small value of 10~$\mu$T ($\sim$0.05\% of $\mu_o H_c$ at $\theta = 45\mathrm{^\circ}$) and allowing the magnetization to evolve in time according to the torque felt by each spin, before the field is stepped further.
As the steady state is approached at each field value, all torque values will tend towards zero.
Reaching this state can take a long time without changing the end result, so we wait until the maximum torque in the entire simulation space reduces to below 4~mT.
This specific value was determined by evaluation of the torque profiles of single and coupled islands driven through reversal and chosen so that relatively slow chains of island reversals are allowed to complete.
It is $\sim$100$\times$ smaller than the maximum value seen during reversal of a single island; in coupled arrays, the torque can peak even higher, at $\sim$1~T.
The torque value is assessed every 20~ps, giving a maximum equivalent external field sweep rate of 500~kT$\,$s$^{-1}$.
For all $\alpha$ angles, we rotate the structure to maintain alignment with the simulation grid to minimise staircase effects.

For the PD and SW-PD models, we use the flatspin package.
For all flatspin simulations, we evolve the magnetization during a staircase field sweep by reversing all islands with energies exceeding the appropriate barrier at each field step (either 1~$\mu$T or 10~$\mu$T).
To allow the interactions between islands to be observed, the new dipolar fields are then calculated and the process repeated until no further islands reverse, before further stepping the external field.
All islands that reverse at a single field form a single cascade in which the order of reversal is an analogue of time in the real system, which we refer to as pseudo-time.

For all structures, the external field was applied along the respective anisotropy axis of each array.
A strong coupling regime exists in all simulated structures other than the PD ones of the pinwheel geometry, and we determine the anisotropy axes from the angle of peak $H_c(\theta)$ in the angular range of the regime, as demonstrated in Figure~\ref{fig:pw_unit_astroid} for a single pinwheel unit.
The anisotropy axis identified in the strong coupling regime reflects that seen \emph{exactly} at the point of reversal.
As the domains or strings that mediate the reversal form and propagate, the local configuration of islands and magnetization at the interface between regions that have reversed and have yet to reverse will change.
This will be especially true for very large arrays and, as a result, the alignment of the axis may change slightly; in effect, the array cannot be characterised by a single anisotropy contribution.
There are multiple possible metrics to track this and we select the component of the net array magnetization lying perpendicular to the geometry axis which, in essence, marks the angle where the array is equally likely to reverse with a clockwise or anticlockwise sense of rotation.

The evolution of the net magnetization in a single array reversal from simulations using the SW-PD model is shown in Figure~\ref{fig:sw_sqo_pw_h_nets}(a) and \ref{fig:sw_sqo_pw_h_nets}(b) at a selection of applied field angles for the open square and pinwheel arrays, respectively.
In this dataset, the field step of 1~$\mu$T was used in order to optimise resolution; in all other data, a larger field step value of 10~$\mu$T was used for improved efficiency.
The thick black lines characterise the reversal paths with the field applied at the anisotropy axis, which show the minimum deviation from zero and are the shortest in length, as expected.
The results for the pinwheel structure was -0.86$\mathrm{^\circ}$ for the SW-PD model.
For the PD model, the equivalent value was -0.23$\mathrm{^\circ}$.
Both PW values are of the opposite sign and much less in magnitude than the MM value of +3.0$\mathrm{^\circ}$.
For square arrays, the anisotropy axes were aligned with the geometrical ones in all models, as expected from symmetry.

The misalignments of the anisotropy axes identified above may seem small, but they are characteristic of an important property of the arrays: interactions between the islands.
For the square array shown in Figure~\ref{fig:sw_sqo_pw_h_nets}(a), the reversal paths are symmetric with respect to the sign of the applied field axis, as expected from a structure with a mirror plane along the geometrical axis.
In contrast, the same parameters plotted for the same angles in Figure~\ref{fig:sw_sqo_pw_h_nets}(b) for the pinwheel array are highly asymmetric across the entire angular range explored, due to the reduced symmetry of the array. 
This effect is likely to be even stronger in real systems where the coupling is stronger than that produced in the SW-PD model.

The ideal reversal paths with the field applied at the respective anisotropy axes are shown in \textbf{Figure~\ref{fig:ideal_reversal_model_maps}} for the different arrays (in columns) and for different models (in rows).
The annotations show the number of cascades, $N_c$, and the field range for complete reversal, $\Delta \mu_o H$.
Videos showing evolution of the microstates during reversal are included in Supporting Information.
For the pinwheel array, reversal of each subarray in the PD model [Figure~\ref{fig:ideal_reversal_model_maps}(a)] is nucleated in different corners and largely independent domains propagate towards the centre over 4 cascades.
The SW-PD model [Figure~\ref{fig:ideal_reversal_model_maps}(d)] greatly improves upon this, with the nucleation occurring in all four corners and 2-D domains propagating across both sub-lattices together (see also the videos in Supplemental Information).
This occurs in a single cascade, replicating the pseudo-exchange effect where reversal of the first island(s) trigger the next one(s), as in a chain of dominoes.
The reversal pattern is approximately a mirror image of the that seen in the reference MM model [Figure~\ref{fig:ideal_reversal_model_maps}(g)] as a result of the different anisotropy axis in the SW-PD model.
These properties of homogeneous 2-D reversal in the pinwheel structure are precisely those which make the pinwheel system an interesting one for potential applications and for the study of fundamental models of magnetism, statistical mechanics, and critical phenomena.

For the open (middle column) and closed (right column) square arrays, the exact reversal is different between the PD and SW-PD models, but they share the same general features of 1-D Dirac strings nucleating near the edges and extending in a direction parallel to the applied field axis, in a number of cascades ranging from 2 to 5.
In the closed square array, the outer ring of islands is one of the first chains to reverse.
This and the other features discussed are also present in the MM simulations [Figure~\ref{fig:ideal_reversal_model_maps}(h) and \ref{fig:ideal_reversal_model_maps}(i)], but the biggest difference is that the MM models produce reversals that are more disordered and thus spread over many more cascades (10--14) and a larger field range (similar results were obtained with the field offset from the geometrical axis by 2$\mathrm{^\circ}$).
These differences are due to the presence of end-states in the MM model, which adds another dimension to the microstate.
The end-states are relatively sensitive to the local field arrangements and are easily modified by the dipolar fields from island reversals.
As the end-states themselves mediate the reversal of each island, the added microstate dimension imparts a degree of apparent disorder in the reversal path.
Although this property can be tailored through choice of the island shape and material, it is intrinsic to the geometry and only present in the MM model where magnetization within the island itself is accounted for.
In connected kagome systems, a related contribution to intrinsic disorder has also been observed to potentially arise from different types of domain walls which mediate reversal in that system.~\cite{Ladak_2012_njp_honeycomb}

In the next section, we further explore the influence of disorder, both intrinsic and extrinsic, in the different models for square and pinwheel arrays.

\section{Reversal of Disordered Arrays}
\label{sec:disorder}
Disorder in ASI may arise from a number of mechanisms, and can be separated into physical properties and dipolar interactions.~\cite{Fraleigh_PRB_17, Kempinger_PRR_2020}
The former include variance in the material volume and moment, in the island shape and rotation, in its position within the array, and from edge roughness,~\cite{Gadbois1995_transMag_edge_roughness, Kohli2011_PRB_edge_roughness_SQ, Velo2020_kagome_micromag} while the later results from stochastic inter-island interactions and are strongly influenced by the geometry of the array.
In the square lattice, it has been shown that several different sources of disorder have a similar effect on the properties of the array,~\cite{Budrikis2012_jap_disorder_sq} and that disorder can be designed into a system to enable access to the ice-rule phase when the sub-lattices lie on separate planes.~\cite{Di_Pietro_Mart_nez_2020_disorder}

While detailed procedures have been developed for the assessment of disorder in coupled systems,~\cite{Hellwig_2007_APL_disorder, Fraleigh_PRB_17, Kempinger_PRR_2020} it is commonly estimated using collective properties such as avalanche critical exponents~\cite{Mengotti_NatPhys_2011} and M-H loops~\cite{Ladak_2010_natPhys_monopole_kagome_nphys1628, Daunheimer_PRL_2011_kagome_disorder} in the kagome lattice; from correlations~\cite{Pollard2012_PRB_charge_dirac_disorder} and vertex populations~\cite{budrikis_PRL_2012_sq_disorder, Pollard2012_PRB_charge_dirac_disorder} in the square one; and from mesoscale magnetic texture during reversal in pinwheel arrays.~\cite{Jensen_2020_flatspin}

In addition to considering both sources of disorder, it is also important to consider the appropriateness of the assumptions of any model used to compare against experimental data.
Indeed, recent work in modelling the kagome system with micromagnetics has raised questions over the attribution of the source and magnitude of disorder from Ising models.~\cite{Velo2020_kagome_micromag}
The effect of ordering and of the non-Ising nature of the same islands in the system investigated in this work is quite different; the AFM ordering in the square geometry serves to promote disorder, while the FM ordering in pinwheel ice suppresses its influence.
Consequently, the appropriateness of each model varies in this system, and so we deal with each geometry differently here.
First, we consider briefly the role of end-state induced disorder in the square geometry before investigating how well the the effect of disorder is approximated for the different models applied to the pinwheel geometry.

Disorder is included in our MuMax3 micromagnetic calculations through the addition of random static fields drawn from a Normal distribution, applied to each island along their long axis.
For the PD and SW-PD models, we use the mechanism built into the flatspin package, whereby the coercive fields or astroid barrier parameter, $H_k$, follows a Normal distribution.
In both models, the disorder is fixed across each reversal and we label it by its standard deviation, $r$, as a fraction (or percentage) of the coercive field (of 20.1~mT) of an isolated island with the field applied at 45$\mathrm{^\circ}$ .
While these disorders correspond to a lateral shift of the astroid in the MM model and to a uniform scaling of the astroid in the SW-PD one, the effects will be similar, especially at the angles around the geometry axes used here (see Supporting Information for a comparison of the two approaches).
The external fields are applied along the respective anisotropy axis of each simulation.

\begin{figure}[hbt]
  \centering
      \includegraphics[width=8.5cm]{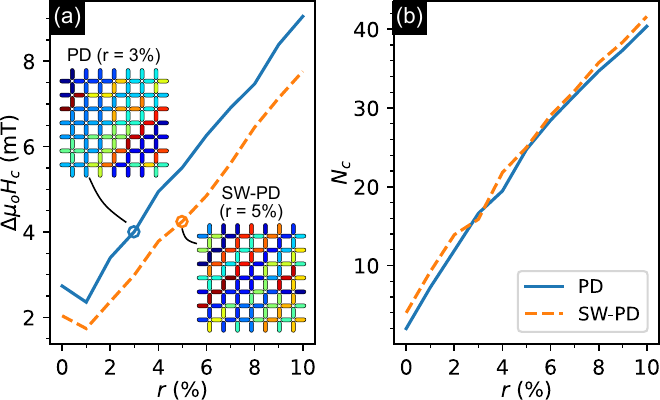}
      \caption{The effect of disorder on (a) the spread in coercive fields and (b) the number of cascades in the PD and SW-PD models applied to an open square array.
      All results were averaged over 32 repeats, with the field applied along the anisotropy axis, and field steps of 10~$\mu$T.
      The insets to (a) show reversal maps for the two models at points that approximate the disorder seen in the MM simulations of Figure~\ref{fig:ideal_reversal_model_maps}(h).}
      \label{fig:sqare_open_disorder_model_params}
\end{figure}

The MM results in Figure~\ref{fig:ideal_reversal_model_maps} highlight the fact that end-states in realistic square systems impart a degree of disorder that is naturally entirely absent in the PD and SW-PD models.
To investigate the strength of this effect, we performed 32 repeats of PD and SW-PD simulations of the open edged square array with different degrees of disorder.
\textbf{Figure~\ref{fig:sqare_open_disorder_model_params}(a)} and \textbf{\ref{fig:sqare_open_disorder_model_params}(b)} show the average spread in coercive field and the number of cascades as a function of disorder, respectively.
The number of cascades [Figure~\ref{fig:sqare_open_disorder_model_params}(b)] is largely independent of the model used, but the spread in coercive fields is offset to lower values in the SW-PD model [Figure~\ref{fig:sqare_open_disorder_model_params}(a)].
This effect is most likely a result of the lower reversal barrier for dipolar field components lying away from the external field direction in the SW-PD model.
Interestingly, both the PD and SW-PD coercive field spread curves show a small negative gradient at very low disorder.
This is due to a reduction in average `jamming' as a result of washing out edge effects,~\cite{budrikis_PRL_2012_sq_disorder} and enabling a greater sampling of all reversal paths.~\cite{budrikis2011_prl_sq_disorder}

Through comparison with the MM model results from an ideal array [Figure~\ref{fig:ideal_reversal_model_maps}(h)], we estimate the amount of extrinsic disorder needed to approximate the level intrinsically present in the MM results as 3\% in the the PD model and 5\% in the SW-PD model, for this particular metric.
Examples of the spatial reversal maps at these two points are shown in the inset to Figure~\ref{fig:sqare_open_disorder_model_params}(a) and are similar to the MM model results of Figure~\ref{fig:ideal_reversal_model_maps}(h).
While the simple comparison made here is not a full exploration of the details of the magnetic texture during reversal, it does serve to demonstrate how different models can lead to different estimates of disorder in an otherwise ideal system.
As we will show below, the same is even more true for pinwheel arrays.

Unlike in the square array, the reduced symmetry of the pinwheel array allows there to be a unfrustrated ground state of s-type end-states.~\cite{Paterson_2019_PRB_pinwheel_FM}
It is partly because of this property that the reversal in an ideal pinwheel array occurs through a single cascade spanning the entire system (in the SW-PD model where end-states are absent, it occurs because of first nearest neighbour coupling during reversal, as discussed in the introduction).
How sensitive this property of homogeneous single cascade reversal is to disorder depends on the strength of the interactions between islands.
Thus, while the SW-PD and MM models appear to produce similar results for the ideal sample [\textit{c.f.} Figure~\ref{fig:ideal_reversal_model_maps}(d) and \ref{fig:ideal_reversal_model_maps}(g)], this may not be true when disorder is included.

In pinwheel arrays, the reversal nucleates at specific corners due to reduced symmetry.
In MM simulations with periodic boundary conditions (PBCs), $\mu_oH_c$ increases from 18.8~mT to 24.0~mT, giving a measure of the reversal energy barrier difference at the edges and bulk of the array.
This higher $\mu_oH_c$ value would never be reached in any finite ideal system due to the propagation of the domain wall modifying the local fields; it is precisely this effect that gives rise to the 2-D ferromagnetism in field-driven pinwheel arrays.
However, it does give some measure of the intrinsic energy landscape and thus the level of disorder that would be needed to completely randomise the array reversal.
From the numbers above, the difference in field of 5.2~mT corresponds to 25.9~\% of $\mu_o H_c$ for an isolated island.

\begin{figure}[hbt]
  \centering
      \includegraphics[width=7.0cm]{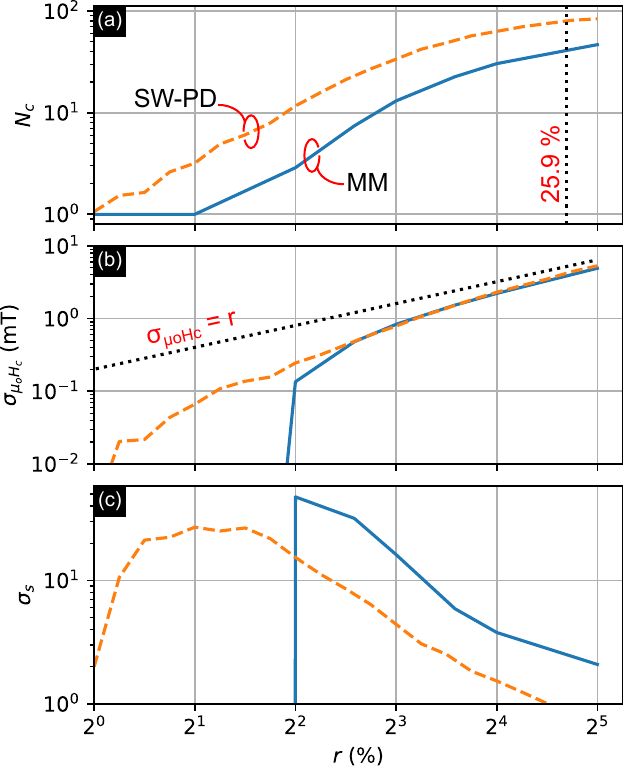}
      \caption{Comparison of three characteristics of field-driven reversal of a 112 island pinwheel array with various amounts of disorder, simulated using the (solid lines) MM and (dashed lines) SW-PD models: (a) the number of cascades, (b) the spread in measured coercive fields, and (c) the spread in the number of islands in all cascades.
      The field step in all simulations was 10~$\mu$T.
      All of the results shown were averaged over 64 repeats, except for the 0\% and 32\% disorder MM ones, which are averaged over 4 and 48 repeats, respectively.
      The black dotted line in (a) shows the degree of disorder equivalent to the difference in coercive field between the edge and bulk of an array in MM simulations.
      The dotted line in (b) marks the line where the measured disorder in coercive fields would match the added disorder.}
      \label{fig:pw_disorder_params}
\end{figure}

To explore the energy landscapes and coupling in the pinwheel array in the SW-PD and MM models, we compare characteristic properties of the reversal as a function of disorder.
\textbf{Figure~\ref{fig:pw_disorder_params}} shows three such parameters as a function of disorder level: the number of cascades across a reversal, $N_c$; the standard deviation of the coercive field, $\sigma_{\mu_oH_c}$; and the standard deviation of the number of islands in a cascade, $\sigma_{s}$, in which the islands need not be contiguous.
For both models, the number of cascades [Figure~\ref{fig:pw_disorder_params}(a)] starts at 1 at zero disorder and, as one might expect, increases towards the system size with increasing disorder.
However, while the number of cascades in the SW-PD model immediately increases to above 1, the MM results stay at 1 until $\sim$2\% disorder.
The higher number of cascades in the SW-PD model remains present across all disorders, indicating that there is weaker coupling between islands in that model than in the MM one. 
In the MM simulations at the 25.9\% disorder identified above, the number of cascades equals approximately one third of the number of islands in the system.

The weaker coupling in the SW-PD model can also be seen in the spread in coercive fields [Figure~\ref{fig:pw_disorder_params}(b)].
The dotted line shows the amount of disorder expected to be measured if the islands were entirely uncoupled.
The SW-PD data and the MM data match well at higher disorder, where the extrinsic disorder is increasingly dominating the island coupling and the curves approach the dotted line.
However, a large difference is seen at lower disorder, where the MM data drops towards zero at disorders of 4\%, while the SW-PD model continues to follow the trend from higher disorders.
In the region of disorder below 4\%, the strong coupling in the MM model largely overcomes the disorder, resulting in all islands reversing at the same field but with a variable reversal path.

The same effect of strong coupling in the MM model can be seen in the spread in cascade sizes across a reversal [Figure~\ref{fig:pw_disorder_params}(c)].
This value should be small at very low and very high disorders, and peak somewhere in between, where a range of cascade sizes is possible.
This parameter follows such a trend in the SW-PD data, but the same threshold disorder in the MM simulations exists (of 4\%), below which variance in the cascade sizes drops to zero.
The larger spread in cascade sizes in the MM model at higher disorder may be due to there being a greater difference in intrinsic island reversal barriers across the finite array in that model due to the edge effect discussed above.

While the SW-PD underestimates the intrinsic disorder in square ASI [Figure~\ref{fig:ideal_reversal_model_maps} and \ref{fig:sqare_open_disorder_model_params}], it overestimates the effect of disorder in the pinwheel geometry.
The range of disorder where the pinwheel geometry is of interest in applications spans a wide range.
In the low disorder range, where the SW-PD model most strongly falls short by not replicating important aspects of the coupling between islands, the use of a dumbbell or multiple-spin model and including spin canting may improve the results, as discussed in the previous section.

\section*{Conclusions}
The ability to accurately model ASI systems is important in order to extract properties of existing systems and to predict those of new ones.
We have investigated three models in the study of field-driven reversals of arrays selected from the square--pinwheel system, exhibiting a range of reversal properties dependant on the island rotation angle.
The two point dipole models with fixed and Stoner-Wohlfarth reversal barriers are many orders of magnitude more efficient than micromagnetic calculations, but do not include as much physics.
Including a Stoner-Wohlfarth barrier when using point dipoles is important for square arrays with disorder, and is critical to approximating the inter-island coupling in pinwheel arrays.

Our results show that some emergent aspects of the coupling between islands in the pinwheel array, such as strong-coupling and anisotropy during reversal, and the influence of disorder are not fully reproduced in existing versions of the Stoner-Wohlfarth point-dipole model.
The inclusion of spin canting and representing each macrospin by a dumbbell or by multiple point dipoles has the potential to improve the modelling of these effects and, in principle, these could easily be included in such a model while maintaining much of its computational efficiency.
Inclusion of these features may also improve the modelling of field-driven demagnetization protocols, which is especially important for pinwheel ice as an experimental ground state has yet to be achieved outside of thermal annealing.

The insights above not only show important differences in existing models of field-driven reversals of ASI systems and how modelling of them can be improved, but also that modifying properties of the islands of an array can have a greater effect than previously appreciated on the emergent properties of the system.
As the array properties fundamentally emerge from those of the constituent islands, we expect many of the results reported will also apply to other ASI systems, but modified according to the array topology.
Whether the discrepancies between results from different models is important for the application of a given structure clearly depends on the particular property or properties of interest.
Nevertheless, the computational efficiency of the simpler models means they will continue to make practical the modelling of larger systems, even if some aspects of the results are different; our work highlights what some of these differences are and the care that must be taken in interpreting them.

\section*{Acknowledgements}
This work was supported by the Leverhulme Trust and the University of Glasgow through LKAS funds.
G.M.M. acknowledges support by the Carnegie Trust for the Universities of Scotland, and the Swiss National Science Foundation under the auspices of Grant No. 200020\_172774.

\section*{Conflict of Interest}
The authors declare no conflicts of interest.


\onecolumngrid

\clearpage
\newpage

\begin{figure*}
\centering
    \textbf{Table of Contents}\\
    \medskip
    \includegraphics[height=50mm]{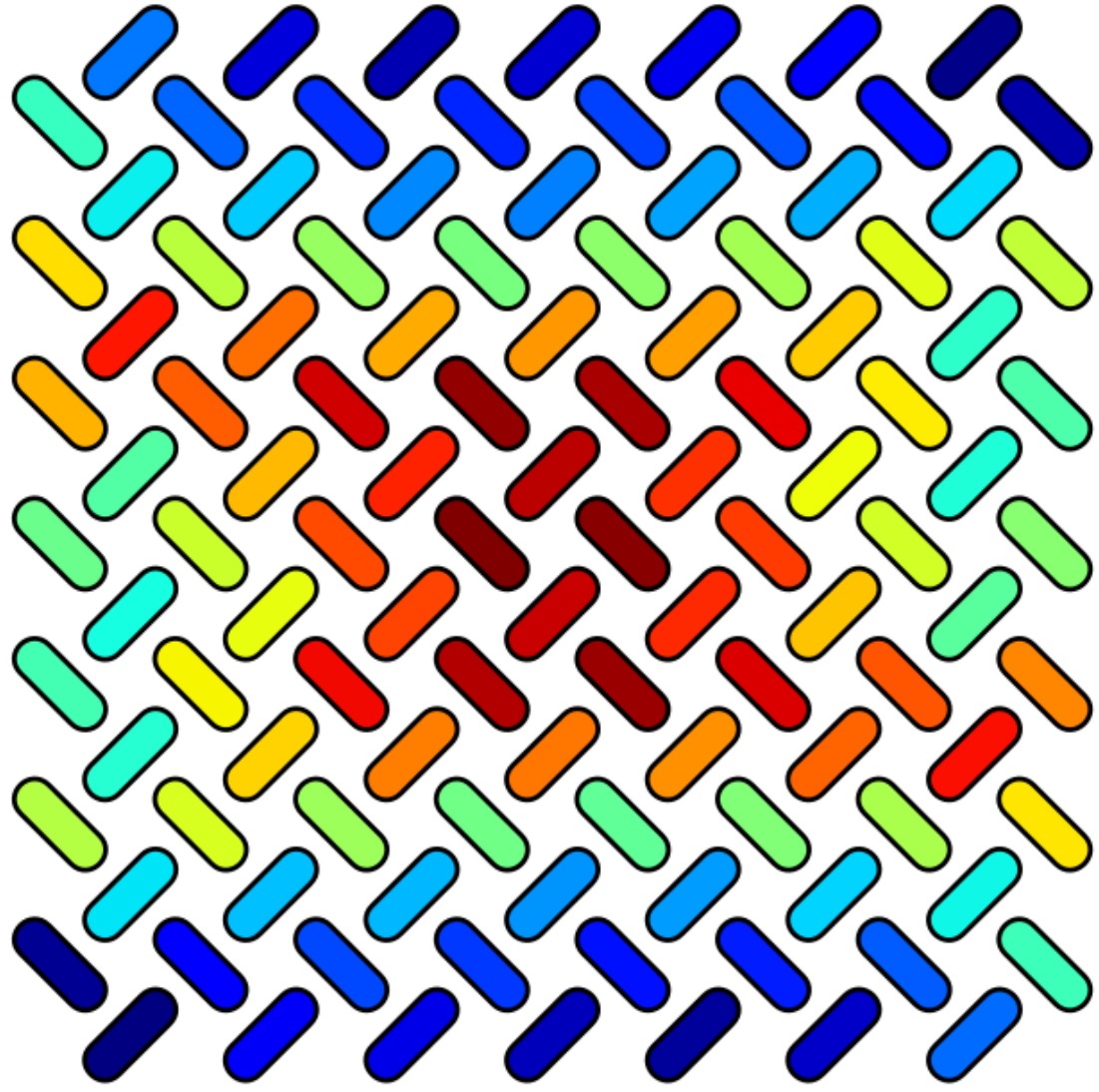}
    \medskip
    
    The collective behaviors of arrangements of ferromagnetic islands in artificial spin ices (ASIs) are of great interest in applications.
    The point dipole approximation has been the mainstay of numerical simulations of ASI systems due to its efficiency.
    By comparing this approach to more expensive micromagnetic modelling of the interactions, we identify important contributions to key aspects of emergent phenomena.
\end{figure*}

\end{document}